\documentclass[12pt]{iopart}
\usepackage{graphicx} 
\usepackage[english]{babel}
\usepackage[utf8]{inputenc}

\begin{document}

\title{Experimental analysis of a physical pendulum with variable suspension point}
\author{Martín Monteiro$^1$, Cecilia Stari$^2$, Cecilia Cabeza$^3$, Arturo C. Mart{\'i}$^3$ }

\address{$^1$ Universidad ORT, Montevideo, Uruguay}
\address{$^2$ Instituto de F\'{i}sica, Facultad de Ingeniería`'',
  Universidad de la Rep\'{u}blica,
 Uruguay}
\address{$^3$ Instituto de F\'{i}sica, Facultad de Ciencias,
  Universidad de la Rep\'{u}blica,
 Uruguay}

\ead{marti@fisica.edu.uy}

\date{\today}
\begin{abstract}
A physical pendulum with variable point of suspension (and, as an outcome, variable inertia moment)
is experimentally analysed. In particular, the period of the small oscillations as a function of position
of the suspension point is measured using three different methods: a smartphone used both as an independent tool or as a data-logger and commercial photo-gate. The experimental results are successfully compared with theoretical calculations based on the addition of inertia moments and the Steiner theorem.
\end{abstract}

\maketitle
\textbf{Physical pendulum}. Concepts about inertia and torque are of paramount importance in almost all introductory Physics courses in high school and introductory university level. In particular, one of the typical examples is the physical pendulum
(also known as compound pendulum) which consists of a rigid body that can freely rotate  around a horizontal axis through a fixed center of suspension. The period of the small oscillations, $T$, depends on the mass $M$,
the distance from the suspension point to the center of mass $R$ and the inertia moment $I$ as
\begin{equation}
    T= 2 \pi \sqrt{\frac{I}{M g R}}
    \label{eq:period}
\end{equation}
where $g$ is the gravitational acceleration. 

In the  experiment proposed here, the period of the small oscillations of a physical pendulum 
whose point of suspension, and then, its inertia moment can be controlled is experimentally analyzed using different modern technologies
\cite{monteiro2014exploring,monteiro2014rotational,patrinopoulos2015angular,salinas2019demonstration}.
As it involves key concepts in classical mechanics and can be readily implemented virtually in any Physics laboratory, the present experiment could encourage students' interest and motivation to experiment by themselves.

\begin{figure}[h]
\begin{center}
\includegraphics[width=0.44\textwidth]{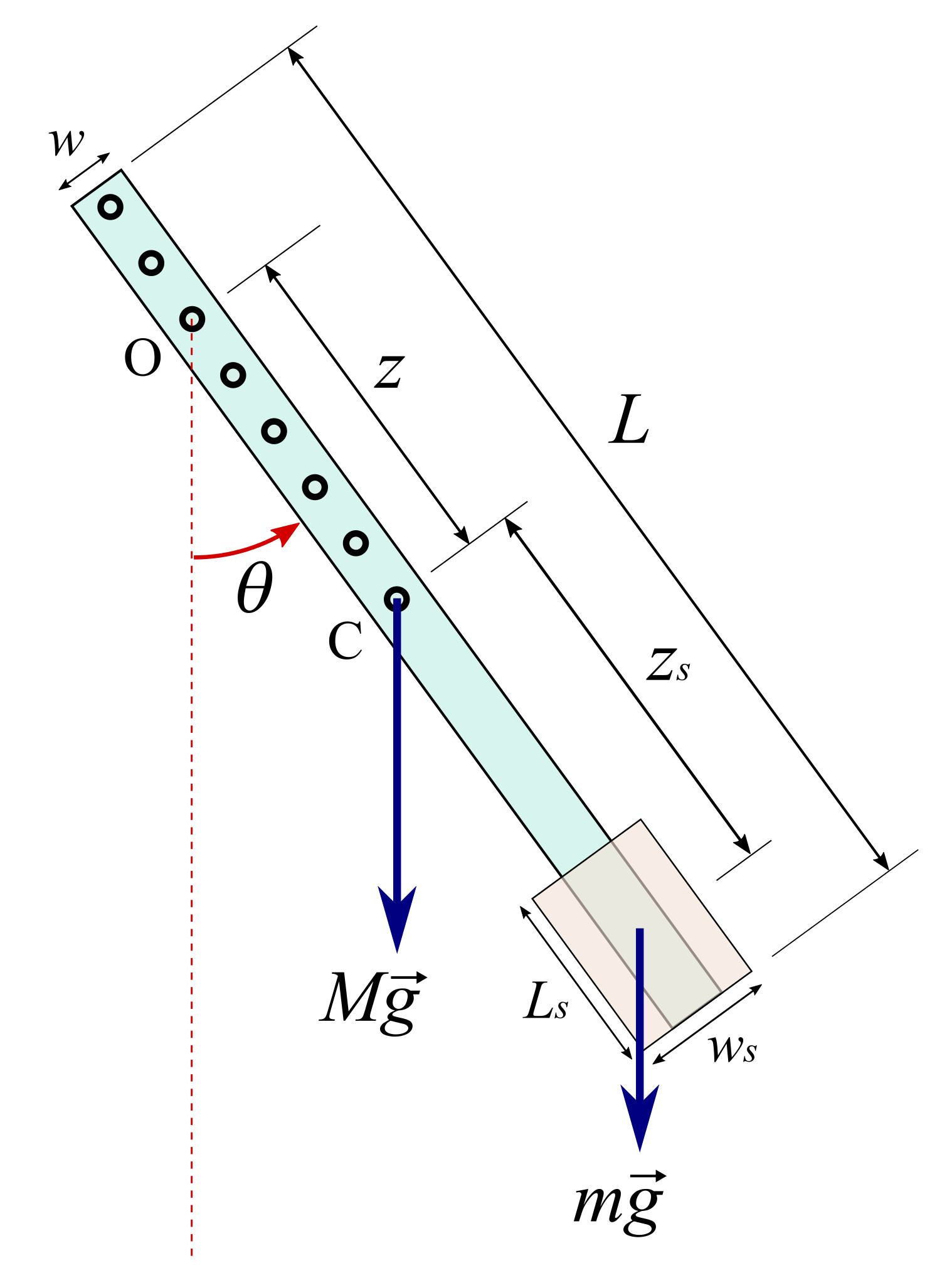}
\caption{Experimental setup consisting of a bar with equispaced suspension points and a smartphone. The labels are described in the text.
}
\label{fig:setup}
\end{center}
\end{figure}

\textbf{Experimental implementation.} The experimental  setup,  depicted in Fig.~\ref{fig:setup}, consists of a rigid metallic bar with equispaced holes and a smartphone. As there are several possible suspension points, the radius of gyration and the inertia moment depend on the selected point. The dimensions of the bar, with holes made at points separated a uniform distance of $1.0$ cm, are $L=1.199$ m  and $w=0.024$ m, and the  mass $M=0.2518$ kg. The smartphone is a Nexus 5, with mass $m=0.1311$ kg, length $L_{s}=0.135$ m and $w_{s}=0.068$ cm.
The distance from the suspension point, $O$, to the center of mass of the bar,  $C$, is indicated with $z$,  while the distance from $C$ to the center of mass of the smartphone is  $z_s$.

Using Eq.~\ref{eq:period} and the geometrical characteristics of the systems the period of the physical pendulum in the regime of small  oscillations can be written as 
\begin{equation}
 T = 2 \pi \sqrt{\frac{I}{\left( \left(M + m \right) z + m z_s \right) g}} 
\end{equation}
where the inertia moment, $I$ is obtained as the sum of the contributions
from the bar and the smartphone $ I = I_{bar} + I_{s} $. Applying the well-known Steiner theorem
the inertia moments can be expressed as 
\begin{equation}
  I_{bar} = \frac{1}{12} M \left( L^2 + w^2 \right) + M z^2   
\end{equation}
and
\begin{equation}
I_{s} = \frac{1}{12} m \left( L_{s}^2 + w_{s}^2 \right) + m \left( z + z_{s} \right)^2
\end{equation}

\textbf{Methods and analysis.} The period of physical pendulum was measured by three different experimental methods:
\begin{itemize}
    \item \textbf{Phyphox:} direct measurement on the smartphone screen using the \textit{Pendulum} tool of the \textit{Phyphox app}. From a temporal series of the angular velocity (about $30$s)  provided by the gyroscope sensor,     the app automatically calculates the period by means of an autocorrelation function (similar to a  FFT but taking the  correlation between maximums of several cycles). The advantage is that it is a direct measure for which  only the  smartphone with the free app is needed. 

\item \textbf{Vernier:} direct measurements using a LabQuest interface and an optical barrier. The advantage is that direct values are obtained even if there is cell phone mounted on the bar. The disadvantage is that it is necessary to dispose of a commercial interface or some equivalent home-made system.

\item \textbf{Androsensor:} smartphone as a data-logger and data processing on a PC. In this procedure, a temporal series of the gyroscope sensor (z axis) was taken for $10$ s, and saved in the smartphone using the \textit{Androsensor app}. All generated \textit{.csv} files (one for each suspension point) were exported to the PC and analyzed using a simple software package (in this case, \textit{Scilab}) and, by means of a sinusoidal fit the period of oscillation is obtained. This method can be useful for experimental courses in which students must develop data management techniques and implement signal processing algorithms in different languages such as Python or Matlab.
\end{itemize}

\begin{figure}
\begin{center}
\includegraphics[width=0.9\textwidth]{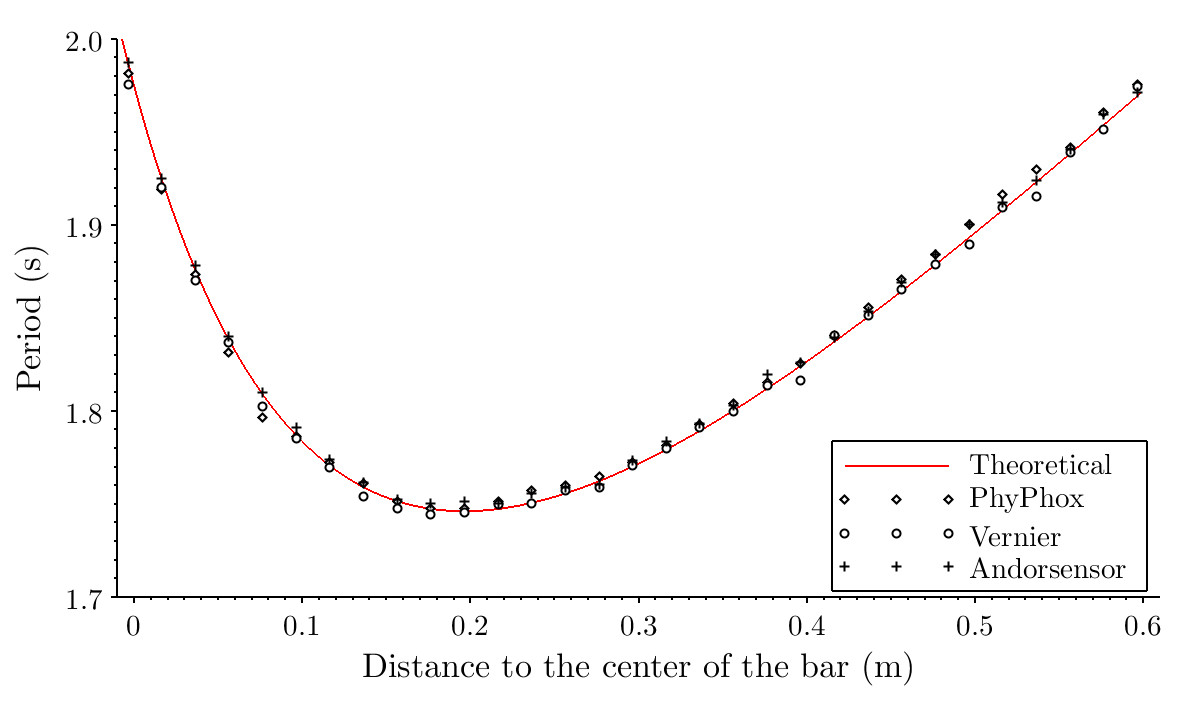}
\caption{Experimental results (symbols) and model (continuous line) for the period of the small oscillations
as a function of the distance of the suspension point to the center of the bar. The root-mean-square deviation
with respect to the theoretical model are: $5.0$ms (Phyphox), $4.0$ms (Vernier) and $3.7$ms (Androsensor). 
}
\label{fig:period}
\end{center}
\end{figure}

Experimental results using the three methods (symbols) and the theoretical prediction (continuous model), plotted in Fig.~\ref{fig:period}, display a great agreement. The root-mean-square deviations with respect to the theoretical model, also indicated in the figure caption, reveal that the three methods, each one with \textit{pros} and \textit{cons} are valid and possible implementations of the present experiment. 
It is also worth noting that the period displays a minimum for a particular distance to the center of the bar. This point could be also the object of an interesting classroom discussion. 
 To sum up, we presented a simple experiment that involves relevant concepts of classical mechanics using moderns technologies that can be readily implemented in a Physics laboratory.

\textbf{Acknowledgements.} We acknowledge financial support from CSIC (UdelaR, Uruguay) and 
Programa de Desarrollo de las Ciencias Basicas (Uruguay).

\section*{References}

\bibliographystyle{abbrv}
\bibliography{/home/arturo/Dropbox/bibtex/mybib}
\end{document}